\documentclass{article}
\usepackage{graphicx}
\usepackage{amsmath}
\usepackage{amsfonts}
\usepackage{amssymb}
\begin{document}
\title{Effect of a Staggered Magnetic Field on the \\
S=1 Haldane Chain with Single-ion Anisotropy}


\author{F.~Anfuso, E.~Ercolessi, G.~Morandi and M.~Roncaglia\\
{\small{{\it Physics Department, University of Bologna, INFM and INFN,}}} \\
{\small{{\it V.le  Bert-Pichat 6/2, I-40127, Bologna, Italy.}}} }

\maketitle


\begin{abstract}

We analyze the gaps in the excitation spectrum of a Haldane chain with single-ion anisotropy in a staggered field.
We show that the gap along the direction of the field increases at a faster rate than the others, while its
spectral weight decreases, being transferred to a two-magnon continuum.

\end{abstract}

\vskip1cm

In recent times the preparation of a new class of magnetic materials of the general formula (R$_x$Y$_{1-x}$)$_2$BaNiO$_{5}$, where R is one of the magnetic rare-earth ions, has arisen a new theoretical interest for the study of integer-spin Haldane chains in a {\it staggered} magnetic field.  The reference compound Y$_{2}$BaNiO$_{5}$ \cite{DR} is one of the best examples of a magnetic material with a singlet ground state and a spin gap which has been interpreted in the framework of the  Haldane conjecture \cite{H}, according to which integer-spin chains have  disordered ground state separated by a gap from a degenerate  triplet of excitations. In all these materials \cite{ZY}, the Ni ions are arranged along linear chains (the a-axis of their orthorhombic structure) with a relatively large in-chain antiferromagnetic coupling ($J\approx 300~ K$) and negligible inter-chain coupling. When partially or totally substituted to the yttrium, the magnetic R ions get positioned between neighboring Ni chains and are weakly coupled to them.  Below a certain N\'{e}el temperature $T_{N}$ (typically: 16 K$\lesssim T_{N}\lesssim80$ K) they order antiferromagnetically. This has the effect of imposing an effective staggered (and commensurate) field on the Ni chains, whose intensity can be indirectly controlled by varying the temperature below $T_{N}$. A phenomenological approach to the study of these compounds has been first presented in \cite{Z}, where the authors propose a simple mean field model, in which the bare magnetization curve for the Ni sublattice is assumed in the form of the staggered magnetization of a single S=1 spin chain, with the effective staggered field proportional to the magnetization of the R ions. The so obtained magnetization curve as function of the magnetic field is in good agreement with the experimental data.

The staggered field is expected to partially lift the degeneracy of the Haldane triplet, leading to different spin gaps in the longitudinal (i.e. parallel to the field) and transverse channels. Surprisingly, up to our knowledge there is only one experiment \cite{R}, analyzing  spin-polarized inelastic neutron scattering on Nd$_2$BaNiO$_{5}$ crystals, which yields a clear evidence for the longitudinal mode.  In this experiment it is found that at $T> T_N=48 ~K$ the value as well as the intensity  of the transverse and the longitudinal modes are practically equal, as in the case of an isolated Haldane chain. To be precise, the resolution of the equipment is high enough to resolve two slightly different peaks in the spectral density, resulting from a very weak single-ion anisotropy along the chain. Below $T_N$, the two excitations clearly separate, with both gaps increasing as the temperature is lowered (the effective staggered field is increased), but with two different rates. Not only the  longitudinal gap has a much slower increase but also its weight seems to be suppressed below $T_N$, contrary to what happens to the transverse mode whose weight is not affected by the magnetic order.  \\

A theoretical approach to  isotropic integer-spin chains in a staggered magnetic field has been first considered in \cite{MZ}, starting from the familiar mapping \cite{H,A} of the Heisenberg model onto a  nonlinear $\sigma$-model  (NL$\sigma$M) effective Lagrangian, that describes the continuum limit, low-energy physics of the Heisenberg chain. What the authors
discuss  very accurately is  actually a related and somewhat more
phenomenological model in which the strict NL$\sigma$M constraint is
softened, then replacing the original NL$\sigma$M with a theory of the
Ginzburg-Landau type parametrized by an adequate set of adjustable parameters. Some recent very accurate numerical studies \cite{Y}, based on an extensive Density Matrix Renormalization Group (DMRG) analysis of an S=1 Heisenberg chain in a staggered field, have revealed a significant discrepancy with the above mentioned theoretical approach, from which it differs in the high field regime and especially as far the longitudinal channel concerns. This has led also the authors of ref. \cite{Y} to question the validity of the NL$\sigma$M approach.

This problem has been solved in \cite{EMPR}, where it has been shown that
an accurate treatment of the NL$\sigma$M does indeed lead to an excellent agreement between the analytical and the  DMRG results. We will review this treatment in some detail below, while applying it to the case in which the Hamiltonian contains also a single-ion anisotropy term. Let us now recall only its main features. In this approach the  NL$\sigma$M  constraint is not softened, but enforced consistently at each level of approximation. In the theory there is then  just a single free parameter, which is assumed to be the (zero-field) Haldane gap and taken from the DMRG data. Also, a consistent scheme of calculation of the propagators and their analytical structure is developed, which shows remarkable  differences between the physical properties of the transverse and the longitudinal channels, already at the tree-level of a loop expansion.  As expected, the transverse propagators  has only two simple poles so that the theory is purely bosonic, with the spectral weight completely exhausted by this magnon excitation. As a consequence, the  transverse gap $\Delta_T$  coincides with the Haldane gap.
On the contrary, the longitudinal propagator has a much richer structure. The dynamical structure factor has again two well-defined poles corresponding to single-particle excitations. However they are now weighted by a prefactor which steadily decreases as the staggered field is increased. The spectral weight which is lost from the poles gets transferred to a multi-particle continuum, which starts from a  two-magnon threshold $\omega > 2 \Delta_T$. Thus we are no longer in the range of applicability of the so-called single mode approximation (SMA) which establishes a relation between the (longitudinal) susceptibility and the gap $\Delta_L$. Once the continuum is taken into account, the agreement between the NL$\sigma$M and the DMRG results become remarkable.
Let us finally mention  that this theoretical scenario has also been proposed \cite{EMR} to explain recent experimental data on CsNiCL$_3$ \cite{Cs} which were not well described by the NL$\sigma$M on isolated spin chains.\\

Keeping in mind the structure of the (R$_x$Y$_{1-x}$)$_2$BaNiO$_{5}$ compounds, our microscopic starting point is the Hamiltonian
\begin{equation} \label{eq:hstagg}
H = \sum_{i} \left\{J \vec{S}_{i}\cdot\vec{S}_{i+1} +  K( S^z_i)^2
-(-1)^ia\vec{H}\cdot \vec{S}_{i} \right\}
\end{equation}
that describes an antiferromagnetic S=1 Heisenberg chain with a single-ion anisotropy term along the chain z-axis, in a staggered magnetic field. We assume $K \ll J$ and at the end we will distinguish between the cases in which the field is parallel and perpendicular to the anisotropy axis.

We  start analyzing this model by
mapping the Hamiltonian (\ref{eq:hstagg}) onto a NL$\sigma$M, by making the well known Haldane ansatz $\vec{S}_i = S \hat{\Omega}_{i}$ with
\begin{equation} \label{eq:mapret}
\hat{\Omega}_{i}=(-1)^i\hat{n}_{i}\sqrt{1-\frac{l^2}{S^2}}+\frac{\vec{l}_{i}}{S}
\end{equation}
$\hat{n}_{i}$   representing the  slowly-varying local staggered magnetization and $\vec{l}_{i}$ the local generator of angular momentum. In the low-energy continuum limit, the model can be described effectively by the action
\begin{eqnarray}
&&S_{eff} =  \int d\mathbf{x} \lbrace
{\cal L}_{\sigma}-S\vec{H}\cdot\vec{n}(\mathbf{x})+
R|\vec{n}(\mathbf{x})\cdot\hat{z}|^2
\rbrace \label{eq:seffanis} \\
&&{\cal L}_{\sigma} = \frac{1}{2gc}(c^2|\partial_{x}\vec{n}|^2
+|\partial_{\tau}\vec{n}|^2) \label{eq:lehaldane}
\end{eqnarray}
where we have set $\mathbf{x}= (\tau,x)$,  $\int d\mathbf{x} = \int_{0}^{L} dx \int_{0}^{\beta}d\tau$ and the constants $c,g,R$ are fixed by the microscopic parameters via the relations $c=2JSa$, $g=\frac{2}{S}$ and  $R = \frac{KS^2}{a}$.

We can now impose the unitary constraint $ |\vec{n}(\mathbf{x})|^2 = 1$ with the aid of a Lagrange multiplier field $\lambda(\mathbf{x})$ by writing the partition function $Z = Tr \{ \exp[-\beta H]\}$ as the path-integral
\begin{equation}
Z = \int [{\cal D} \vec{n}] \left[ \frac{{\cal D} \lambda}{2 \pi} \right]
\exp\left[-S_{eff} -i \int d\mathbf{x} \lambda(\mathbf{x})(|\vec{n}(\mathbf{x})|^2-1) \right] \label{eq:parfun}
\end{equation}

The isotropic case $R=0$ has already been considered in ref. \cite{EMPR}, to which we refer for all the details of the calculations, that can be easily transposed to the case we are interested here.  In this letter  we will outline  only the main steps that lead to the calculations of the physical quantities.

First we need to promote $Z$ to a generating functional $Z[\vec{J}]$ by replacing $S_{eff}$ with
\begin{equation}
S[\vec{J}] =  \int d\mathbf{x}  \lbrace
{\cal L}_{\sigma}-S\vec{J}(\mathbf{x}) \cdot\vec{n}(\mathbf{x})+
R|\vec{n}(\mathbf{x})\cdot\hat{z}|^2
\rbrace \label{eq:sj}
\end{equation}
and setting $\vec{J}(\mathbf{x})= \vec{H}$ only at the end of the calculations. $S[\vec{J}]$ being quadratic, we can proceed to integrate out the fields  $\vec{n}(\mathbf{x})$ obtaining
\begin{equation} \label{eq:zintegratal}
Z[\vec{J}] \propto \int [\frac{D\lambda}{2\pi}]\exp(-S[\lambda,\vec{J}])
\end{equation}
with
\begin{eqnarray} \label{eq:sintegrata}
S[\lambda,\vec{J}]&=& Tr\lbrace\ln(G^{-1}_{11})\rbrace
+\frac{1}{2} Tr\lbrace\ln(G^{-1}_{11}+R)\rbrace\\
&+&i\int
d\mathbf{x}\lambda(\mathbf{x})-\frac{1}{2}S^2\int
d\mathbf{x}d\mathbf{x}'\vec{J}(\mathbf{x})\cdot G(\mathbf{x},\mathbf{x'})
\vec{J}(\mathbf{x'})
\nonumber
\end{eqnarray}
where $Tr[A(\mathbf{x},\mathbf{x'})]$ stands for $ \int d\mathbf{x}
A(\mathbf{x},\mathbf{x}) $. We have denoted with $G(\mathbf{x},\mathbf{x'})$ the  $3\times 3$ matrix operator whose inverse   $G^{-1}(\mathbf{x},\mathbf{x'})$ has components $G_{\alpha\beta}^{-1}(\mathbf{x},\mathbf{x'})$ (for $\alpha, \beta =1,2,3$)  given by
\begin{equation} \label{eq:gdef}
G^{-1}_{\alpha\beta}(\mathbf{x},\mathbf{x'})=
-\frac{\delta_{\alpha\beta}}{gc}[c^2\partial^2_{x}
+\partial^2_{\tau}+2igc\lambda(\mathbf{x})-gcR\delta^{\alpha
3}]\delta(\mathbf{x}-\mathbf{x'})
\end{equation}

Finally we resort to a saddle-point approximation to implement the constraint by fixing $\lambda(\mathbf{x})$ to its mean field value $\lambda^*(\mathbf{x},\vec{J} )$ through the equation
$ \delta S[\lambda,\vec{J}]/ \delta \lambda(\mathbf{x})=0$
which reads
\begin{equation}\label{eq:stazlam2}
1 = 2G_{11}(\mathbf{x},\mathbf{x})+\frac{1}{G^{-1}_{11}(\mathbf{x},\mathbf{x})+R}  +  S^2\int
d\mathbf{y}d\mathbf{y'}\vec{J}(\mathbf{y})\cdot G(\mathbf{y},\mathbf{x})
G(\mathbf{x},\mathbf{y'})\vec{J}(\mathbf{y'})
\end{equation}
In this approximation,  we can  calculate the connected components of the propagators as
\begin{equation} \label{eq:propmft}
P^{\alpha\beta}_c(\mathbf{x},\mathbf{x'})\approx
-\frac{\delta^2S[\lambda^*,\vec{J}]}
{\delta J^{\alpha}(\mathbf{x})\delta J^{\beta}(\mathbf{x'})}
\end{equation}
which are therefore given by
\begin{eqnarray}
P^{\alpha\beta}_{c}(\mathbf{x},\mathbf{x'})&=&\frac{S^2}{2}\delta^{\alpha\beta}
[G_{\alpha\alpha}(\mathbf{x},\mathbf{x'})+G_{\alpha\alpha}
(\mathbf{x'},\mathbf{x})]
+iS^2\int d\mathbf{y}d\mathbf{y'}
[G(\mathbf{x},\mathbf{y})G(\mathbf{y},\mathbf{y'})\nonumber \\
&+& G(\mathbf{y'},\mathbf{y})G(\mathbf{y},\mathbf{x})]_{\alpha\beta}
J^{\alpha}(\mathbf{y})\frac{\delta\lambda(\mathbf{y'})}
{\delta J^{\beta}(\mathbf{x})}\label{eq:dsdj3}
\end{eqnarray}
The functional derivative $\delta\lambda(\mathbf{y'})/
\delta J^{\beta}(\mathbf{x})$ in the right hand side is implicitly determined by the integral equation
\begin{equation} \label{eq:dsdj6}
\int
d\mathbf{y}H(\mathbf{x},\mathbf{y})\biggl(i\frac{\delta\lambda(\mathbf{y})}
{\delta
J^{\alpha}(\mathbf{x'})}\biggr)
=-2S^2G_{\alpha\alpha}
(\mathbf{x},\mathbf{x'})
\int
d\mathbf{y}G_{\alpha\alpha}(\mathbf{x},\mathbf{y})J^{\alpha}(\mathbf{y})
\end{equation}
with
\begin{eqnarray}
H(\mathbf{x},\mathbf{x'})
&=&\biggl(\frac{\delta^2S}{\delta\lambda(\mathbf{x})\delta\lambda(\mathbf{x'})}\biggr)
_{\vec{J},\lambda=\lambda^*}
=4\Gamma(\mathbf{x},\mathbf{x'})+\frac{2G^{-1}_{11}
(\mathbf{x},\mathbf{x})\Gamma(\mathbf{x},
\mathbf{x'})G^{-1}_{11}(\mathbf{x},\mathbf{x})}{(R+G^{-1}_{11}
(\mathbf{x},\mathbf{x}))^2} \nonumber \\
&+&4S^2\int
d\mathbf{y}d\mathbf{y'}\vec{J}(\mathbf{y})
G(\mathbf{y},\mathbf{x'})G(\mathbf{x'},\mathbf{x})
G(\mathbf{x},\mathbf{y'})\vec{J}(\mathbf{y'})
 \label{eq:dsdj5}
\end{eqnarray}
where  $ \Gamma(\mathbf{x},\mathbf{x'})$ is the polarization bubble $ \Gamma(\mathbf{x},\mathbf{x'})\equiv G_{11}(\mathbf{x},\mathbf{x'})G_{11}(\mathbf{x},\mathbf{x'})$.

At the physical point $\vec{J} = \vec{H} $, one finds that the field $\lambda$ is constant. Setting then $c^2 \xi^{-2} = -2i gc \lambda $ the following results can be found.

Going to Fourier transform in $\mathbf{q} = (q, \Omega_n = 2 \pi n / \beta)$, the sums over the frequency can be exactly performed while, in the continuum limit, it is necessary to introduce a cut-off $\Lambda$ in momentum space.  The saddle point condition (\ref{eq:stazlam2}) then reads
\begin{eqnarray} \label{eq:sadconfou}
1&=&\frac{g}{\pi}\int_{0}^{\Lambda}dq\biggl[\frac{1}{\sqrt{q^2+\xi^{-2}}}
\coth(\frac{1}{2}\beta c\sqrt{q^2+\xi^{-2}})\\
&+&\frac{1}{2}\frac{\coth(\frac{1}{2}\beta c\sqrt{q^2+\xi^{-2}+t})}
{\sqrt{q^2+\xi^{-2}+t}}\biggr]
+S^2(H^2_x+H^2_y)\frac{\xi^4g^2}{c^2}+S^2H^2_z\frac{g^2}{c^2(t+\xi^{-2})^2}
\nonumber
\end{eqnarray}
with $t=Rg/c$. As in \cite{EMPR}, we have chosen to fix the only free parameter of our theory, the cut-off $\Lambda$,  by requiring the correlation length $\xi^{-2}$ to coincide at $\vec{H} =0$ with the numerical result that can be obtained through a Density Matrix Renormalization Group analysis \cite{O} of the original Hamiltonian (\ref{eq:hstagg}) at zero field.

Then the magnetization can be easily calculated through the formula
\begin{equation} \label{eq:magnxi1}
\vec{m}=-\frac{1}{\beta L}\frac{\delta S}{\delta \vec{H}}=
\frac{gS^2}{c}\left(\begin{array}{c}
\xi^2H_x \\
\xi^2H_y \\
\frac{1}{\xi^{-2}+t}H_z \end{array} \right)
\end{equation}

Also, assuming translational invariance, eq. (\ref{eq:dsdj5}) can be solved by going to Fourier transform to obtain
\begin{equation} \label{eq:hfouq}
\widetilde{H}(\mathbf{q})=4\widetilde{\Gamma}(\mathbf{q})+2\frac
{G^{-1}_{11}(0)\widetilde{\Gamma}(\mathbf{q})G^{-1}_{11}(0)}
{(R+G^{-1}_{11}(0))^2}+\frac{4}{S^2}\vec{m}\widetilde{G}(\mathbf{q})\vec{m}
\end{equation}
The latter formula can be now inserted in  (\ref{eq:dsdj6}) in order to calculate the Fourier components of the propagators (\ref{eq:dsdj3}).\\

1) Let us first examine the case in which the applied staggered field is parallel to the anisotropy axis: $\vec{H} = H \hat{z}$. In this case  $m_x=m_y=0$, while $m_z$  as a function of $H$ is obtained from (\ref{eq:magnxi1}).

One finds that the transverse susceptibility $\chi^T= \chi_x = \chi_y$ is given by
\begin{equation} \label{eq:chit1}
\chi^T=\frac{m}{H}=\frac{gS^2}{c}\frac{1}{t+\xi^{-2}}
\end{equation}
Also, the transverse propagator$ \widetilde{P}^T = \widetilde{P}^{xx} = \widetilde{P}^{yy}$ is just a free boson propagator:
\begin{equation} \label{eq:propt2}
\widetilde{P}^T_c(\mathbf{q})=S^2\widetilde{G}_{11}(\mathbf{q})
\end{equation}
with
\begin{equation}
\widetilde{G}_{11}(\mathbf{q})=\frac{gc}{[\Omega^2_n+c^2(q^2+\xi^{-2})]}
\label{g11}
\end{equation}
which, when analytically continued to the real axis, has simple poles at $\omega = \pm c \epsilon (q) $, $\epsilon (q)\equiv \sqrt{ q^2 + \xi^{-2}}$. Thus the spectral weight function is simply given by
\begin{equation} \label{eq3new}
Im\widetilde{G}^T_c(q,\omega)=\frac{\pi gcS^2}{2\epsilon(q)}\lbrace\delta(\omega-\epsilon(q))-\delta(\omega+\epsilon(q))\rbrace
\end{equation}
and has therofore the structure required for the applicability of the SMA \cite{HB}. In this approximation, $ \chi^T = Sgc/(\Delta_T^2)$ with the transverse gap coinciding with the Haldane gap:
\begin{equation} \label{eq4new}
\Delta_T=\Delta_0=c\xi^{-1}
\end{equation}

On the contrary, the longitudinal susceptibility $\chi^L= \chi_z $ has to be calculated as
\begin{equation} \label{eq:chil1}
\chi^L=\frac{dm}{dH}= \chi^T\biggl(1+2H\frac{\xi^{-3}}{t+\xi^{-2}}\frac{d\xi}{dH}\biggr)
\end{equation}
with $\frac{d\xi}{dH}$ obtained from eq. (\ref{eq:sadconfou}):
\begin{eqnarray} \label{eq:dxidh}
\frac{d\xi}{dH}&=&-\frac{2HS^2g^2}{c^2(t+\xi^{-2})^2}\biggl[\frac{g}{\pi}
\frac{\Lambda}{\sqrt{1+(\Lambda\xi)^2}}\\
&+&\frac{g}{2\pi}\frac{\Lambda\xi^{-3}}
{(\xi^{-2}+t)(\xi^{-2}+\Lambda^{2}+t)^{\frac{1}{2}}}
+\frac{4S^2g^2H^2}{c^2}\frac{\xi^{-3}}{(t+\xi^{-2})^3}\biggr]^{-1}
\nonumber
\end{eqnarray}
In this case, the longitudinal propagator $ \widetilde{P}^L = \widetilde{P}^{zz}$ is instead given by
\begin{equation} \label{eq:propl2}
\widetilde{P}^L_c(\mathbf{q})=\frac{S^2(2+W^2)\widetilde{G}_{33}(\mathbf{q})
\widetilde{\Gamma}(\mathbf{q})}
{(2+W^2)\widetilde{\Gamma}(\mathbf{q})
+\frac{2m^2}{S^2}\widetilde{G}_{33}(\mathbf{q})}
\end{equation}
with
\begin{equation} \label{eq:gw}
\begin{aligned}
\widetilde{G}_{33}(\mathbf{q})&\equiv\frac{gc}{[\Omega^2_n+c^2(q^2+\xi^{-2}+t)]}\\
W^2&\equiv \frac{G^{-2}_{11}(0)}{(R+G^{-1}_{11}(0))^2}
\end{aligned}
\end{equation}
Following \cite{EMPR}, we can perform an analytic continuation in the variable $\omega$ ($\omega \rightarrow  z =  \omega + i \eta$) and define:
\begin{equation} \label{eq5new}
\begin{aligned}
G(z,q) =\frac{\widetilde{G}^L_c(q,\Omega_n)}{gcS^2} \; \; &, \; \;
\Gamma(z,q) =\frac{(2+W^2)\widetilde{\Gamma}(q,\Omega_n)}{2gc} \\
\delta =\biggl(\frac{m^2}{S^2}\biggr) \; \; &, \; \;
\varepsilon(q) =\sqrt{c^2(q^2+\xi^{-2}+t)}
\end{aligned}
\end{equation}
Thus we are led to analyze the structure of the function (omitting the specification of the dependence of the label $q$)
\begin{equation} \label{eq6new}
G(z)=\frac{\Gamma(z)}{\Gamma(z)(\varepsilon^2-z^2)+\delta}
\end{equation}
for $ 0\leq \delta \leq 1$ and $\varepsilon \leq 2 \Delta_0$.

Going to the real axis from  above ( $\eta  \rightarrow0^+$), a long but standard calculation shows that, for $|\omega| < 2 \Delta_0$,  the imaginary part of $G(z)$ is given by
\begin{equation} \label{img}
ImG(\omega)=\gamma\frac{\pi}{2\varepsilon_{L}%
}\{\delta(\omega-\epsilon_{L})-\delta(\omega+\epsilon_{L})\}
\end{equation}
so that the longitudinal propagators has simple poles on the real axis at the points $\omega = \pm \epsilon_L$, with $\epsilon_L$ fixed by the equation
\begin{equation} \label{el}
\epsilon_L ^2 = \varepsilon(q)^2  + \frac{\delta}{\Gamma_1(\epsilon_L ^2)}
\end{equation}
where $\Gamma(z)= \Gamma_1(z)+ i\Gamma_2(z)$.
The prefactor $\gamma$ in eq. (\ref{img}) gives the reduction of the quasiparticle weight with respect to the pure bosonic case and is given by
\begin{equation}
\gamma=\left\{ \left[1+\frac{\delta}{\Gamma_{1}^{2}(\omega^{2})}\frac{d\Gamma_{1}%
(\omega^{2})}{d\omega^{2}}\right]_{\omega=\epsilon_{L}}\right\}^{-1}
\end{equation}
The coefficient $\gamma$ is steadily decreasing with increasing $H$, going to the value $1$ quadratically for  $H\rightarrow 0$.

In the range $|\omega| > 2 \Delta_0$, one instead finds:%
\begin{equation} \label{two}
Im G(\omega)=\frac{\delta\Gamma_{2}(\omega)}{(\varepsilon^{2}-\omega^{2}%
)^{2}(\Gamma_{2}(\omega))^{2}+(\delta+(\varepsilon^{2}-\omega^{2})\Gamma
_{1}(\omega))^{2}}%
\end{equation}
As $H\rightarrow  0$, $Im G(\omega)$  vanishes and we  recover  the simple pole structure for both the longitudinal and the transverse propagators.
The longitudinal pole will also survive up to saturation, but with a strongly field-dependent strength:
as the field increases the spectral weight that is lost from the pole gets transferred to
the two-magnon continuum $(\ref{two})$, as it is dictated  by the sum rule \cite{EMPR}:
\begin{equation}
\int\limits_{-\infty}^{+\infty}\frac{d\omega}{\pi}\omega\operatorname{Im}%
G(\omega)=1
\end{equation}

As in the transverse case, we can define the
longitudinal gap $\Delta_{L\text{ }}$ via $\Delta_{L\text{ }}=\epsilon
_{L}(q=0)$, thus obtaining
\begin{equation}
\Delta_{L}^{2}=\Delta_{T}^{2}+ c^2 t + \frac{\delta}{\Gamma_{1}(0,\Delta_{L})}%
\end{equation}
with
\begin{equation} \label{eq7new}
\Gamma_1(0,\Delta_L)=\frac{g(2+W^2)}{4}\int^{+\infty}_{2\Delta_T}\frac{d\omega}{\pi}\frac{1}{\sqrt{\omega^2-4\Delta^2_T}}\frac{1}{\omega^2-\Delta^2_L}
\end{equation}
Due to the presence in the Hamiltonian (\ref{eq:hstagg}) of the single-ion anisotropy term,
the longitudinal and the transverse gaps do not coincide even at $\vec{H} =0$, the former being larger
(smaller) than the latter for $K>0$ ($K<0$). The case $K=-0.02$ and $K=+0.02$ are shown in Fig.1(a) and Fig.1(b)
respectively, where we give $\Delta_L= \Delta_z$ and $\Delta_T= \Delta_x = \Delta_y$ as function of the magnetic field.

\begin{figure}[t]
\centerline{\includegraphics[width=5.9cm]{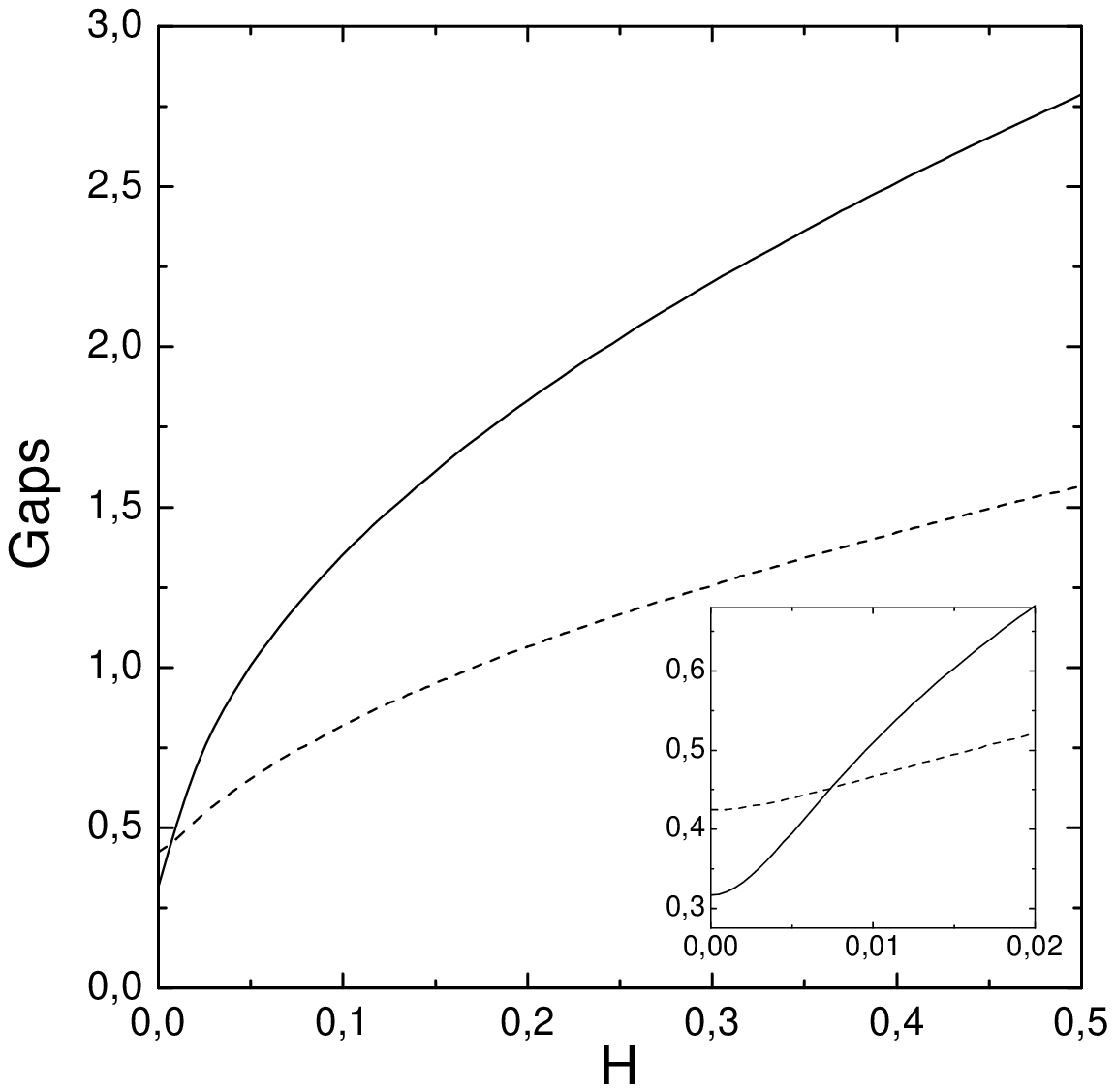}\qquad\quad
\includegraphics[width=5.9cm]{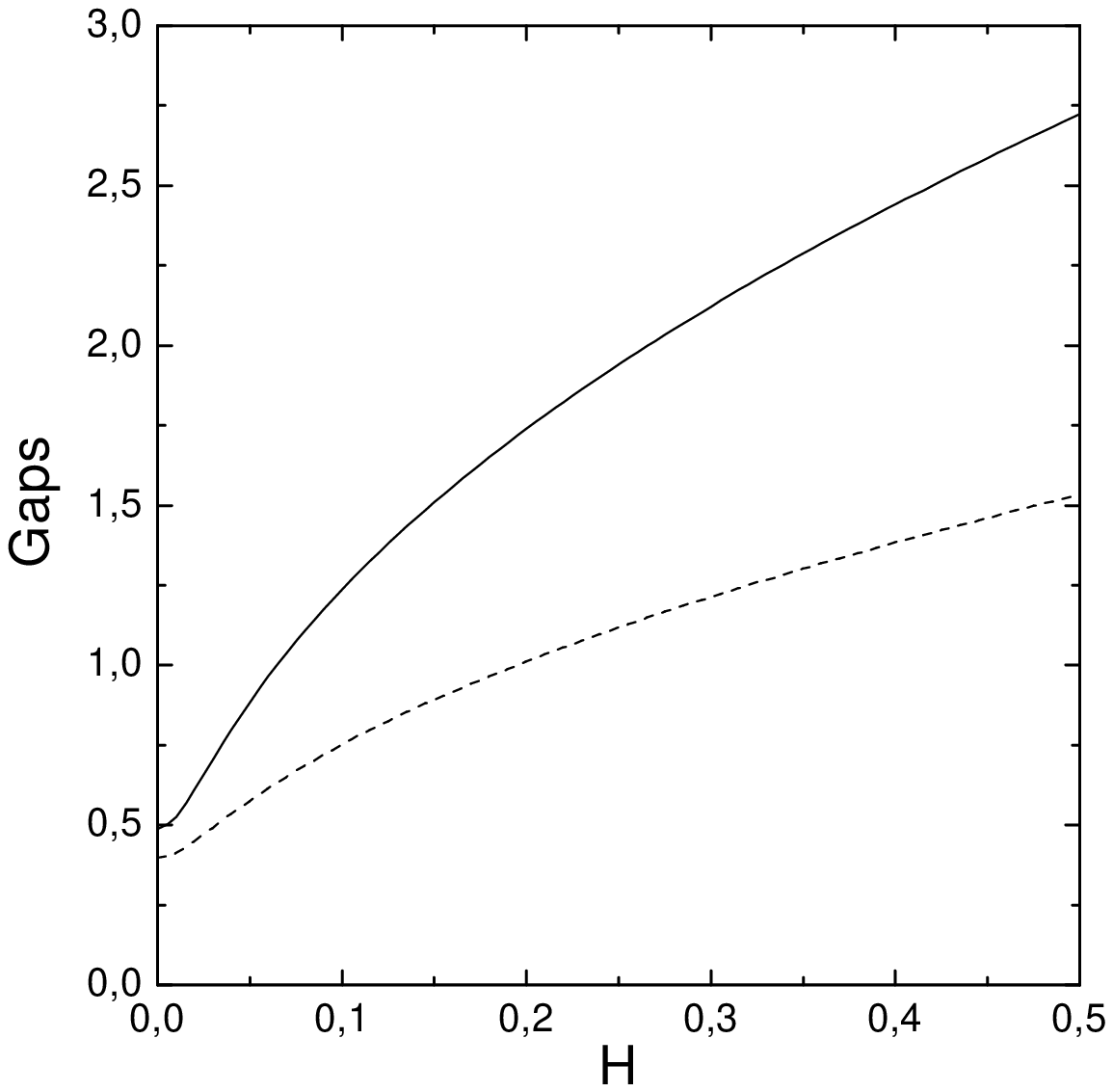}} \caption{The
longitudinal gap $\Delta_L $ (solid line) and the transverse gap
$\Delta_T $ (dashed line) as a function of $\hat{H}= H \hat{z}$
for (a) $K=-0.02$ and (b) $K=+0.02$.}
\end{figure}

We notice that for positive $K$ the longitudinal gap is always higher than the transverse one.
On the contrary, for negative $K$, the longitudinal gap is lower than the transverse one at zero field.
As a function of $H$ the former
increases at a faster rate than the latter, thus resulting in a level crossing at $H \sim 0.075$,
as it is shown in the inset of Fig.1(a).\\

2) Now we examine the case in which the applied staggered field is perpendicular to the anisotropy axis: $\vec{H} = H \hat{x}$.
We concentrate our attention on the structure of the propagators, which now are all different:
\begin{eqnarray}
\widetilde{P}^{xx}_c(\mathbf{q}) &=&\frac{S^2(2+W^2)\widetilde{G}_{11}(\mathbf{q})
\widetilde{\Gamma}(\mathbf{q})}
{(2+W^2)\widetilde{\Gamma}(\mathbf{q})
+\frac{2m^2}{S^2}\widetilde{G}_{11}(\mathbf{q})} \label{eq:propx} \\
\widetilde{P}^{yy}_c(\mathbf{q})&=& S^2 \widetilde{G}_{11}(\mathbf{q})
 \label{eq:propy} \\
\widetilde{P}^{zz}_c(\mathbf{q})&=& S^2 \widetilde{G}_{33}(\mathbf{q})
 \label{eq:propz}
\end{eqnarray}
where $ \widetilde{G}_{11}(\mathbf{q})$, $\widetilde{G}_{33}(\mathbf{q})$ and $W$ are given in
(\ref{g11}) and (\ref{eq:gw}).

For the transverse channels $y,z$ one immediately sees that the propagators are again those of a free bosonic
theory to which one can apply the SMA. Thus the gaps are simply given by
\begin{eqnarray}
\Delta_y &=&\Delta_0 =c\xi^{-1} \label{gapy} \\
\Delta_z &=& \sqrt{\Delta_0^2 + c^2 t}  \label{gapz}
\end{eqnarray}

As before, the SMA is no longer applicable in the longitudinal $x$-direction. We find:
\begin{equation}
\Delta_{x}^{2}=\Delta_{y}^{2}+ \frac{\delta}{\Gamma_{1}(0,\Delta_{x})}%
\end{equation}
with
\begin{equation}
\Gamma_1(0,\Delta_x)=\frac{g(2+W^2)}{4}\int^{+\infty}_{2\Delta_T}\frac{d\omega}{\pi}\frac{1}{\sqrt{\omega^2-4\Delta^2_0}}\frac{1}{\omega^2-\Delta^2_x}
\end{equation}

\begin{figure}[t]
\centerline{\includegraphics[width=6cm]{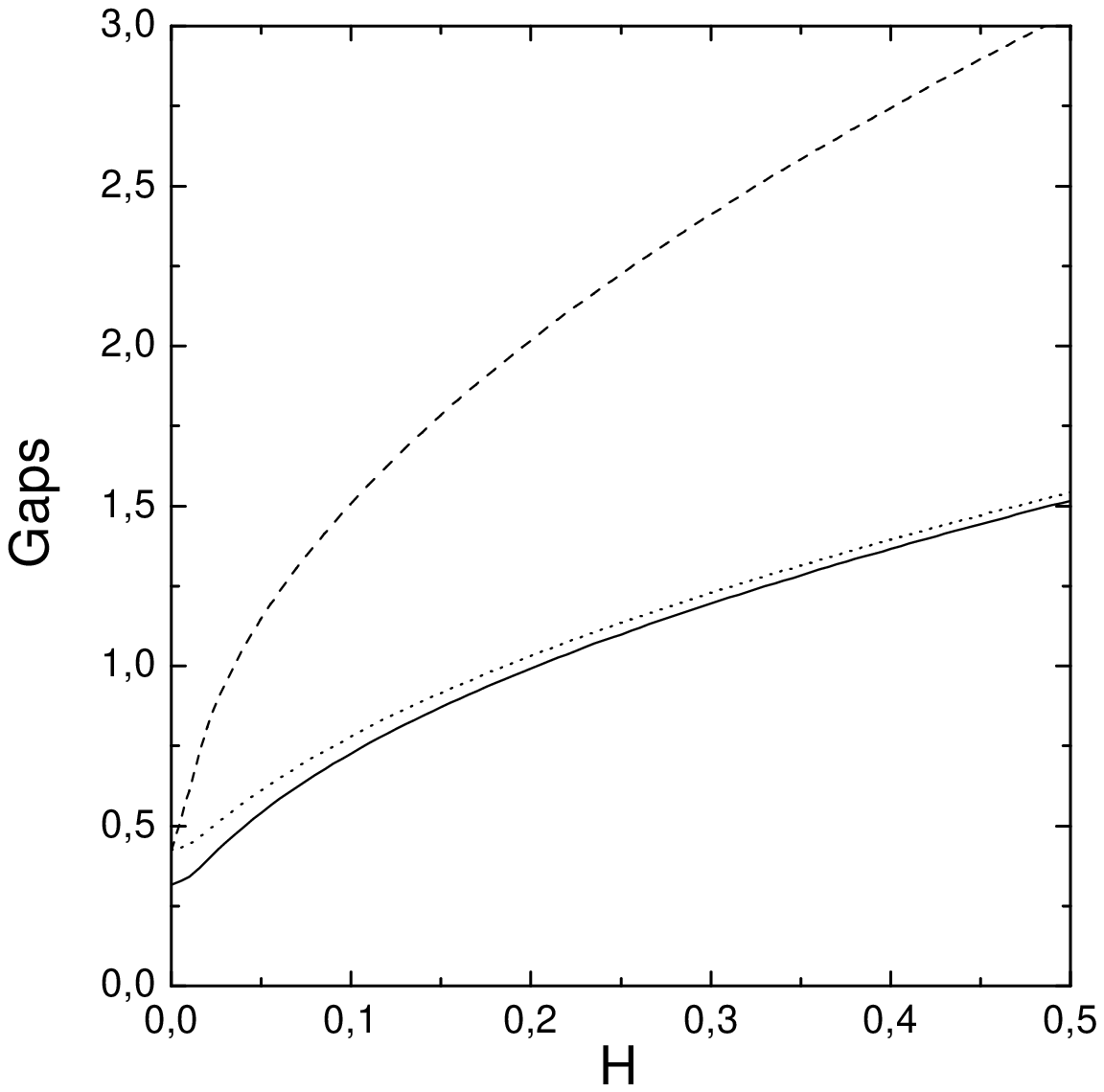}\qquad\quad
\includegraphics[width=6cm]{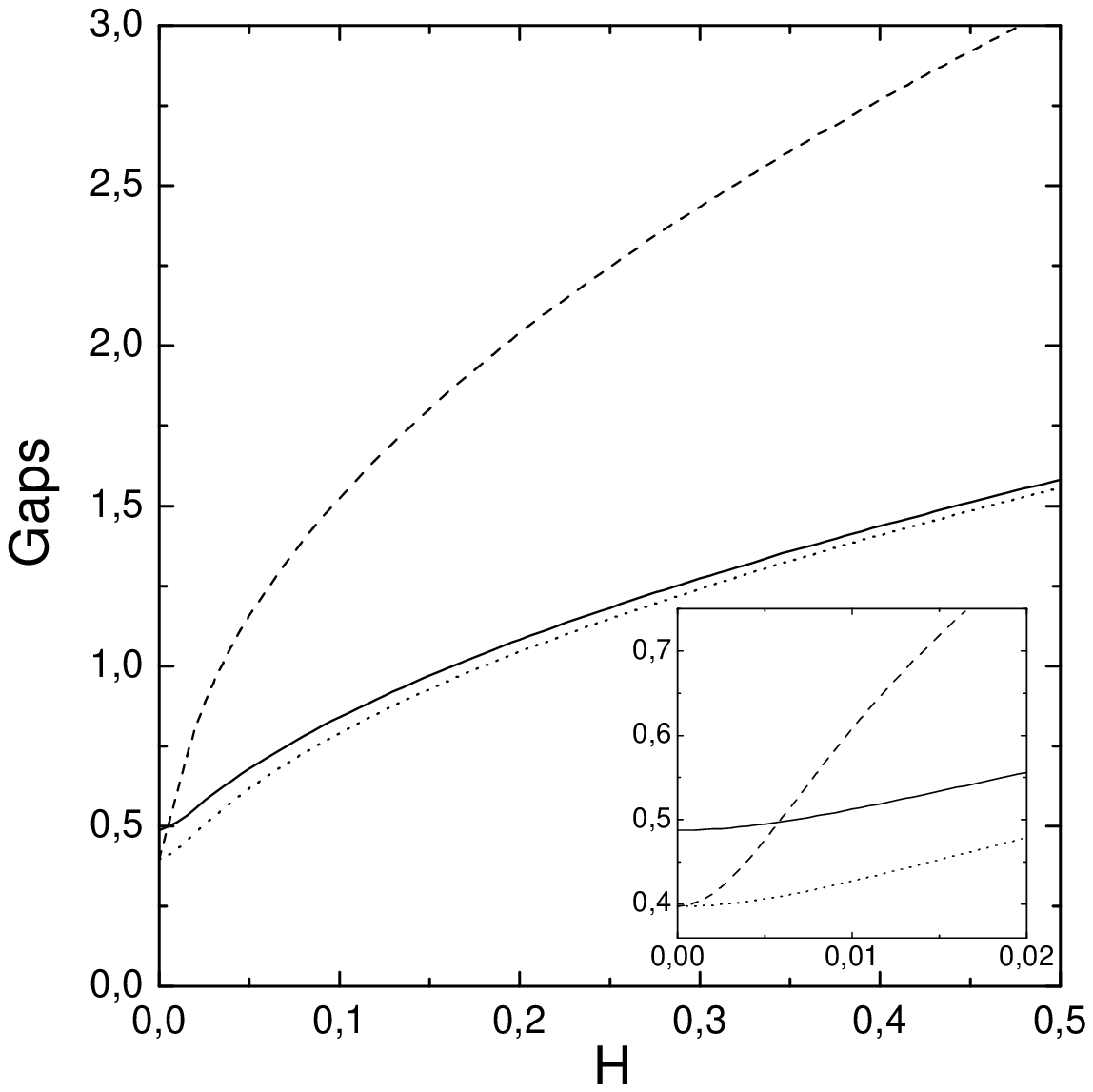}}
\caption{The gaps $\Delta_x $ (dashed line), $\Delta_y $ (dotted line) and $\Delta_z $ (solid line)
as a function of $\hat{H}= H \hat{x}$ for (a) $K=-0.02$ and (b) $K=+0.02$.}
\end{figure}

For negative $K$, we find an initial splitting between a lower gap in the $z$ direction and a higher one in the
$x,y$ directions. As the field is turned on along the $x$-axis, we find a further splitting of $\Delta_x$
and $\Delta_y$, with the former increasing faster. This is shown in Fig.2(a).
On the contrary, for positive $K$ the  splitting at zero field is between a lower gap in the $x,y$ directions and a
higher one in the $z$ direction (Fig.2(b)).
As before, the gap $\Delta_x$  increases faster than $\Delta_y$ and $\Delta_z$ when the field
is turned on along the $x$-axis.
At $H\sim 0.06$ there is a crossing of levels with the gap $\Delta_x$ becoming larger than $\Delta_z$ as shown in
the inset of Fig.2(b).
In both cases, towards saturation the longitudinal gap $\Delta_x$ goes to a value about
twice that of the transverse modes $\Delta_y$, $\Delta_z$, which become approximately equal.
In addition, though surviving up to saturation, the longitudinal mode has a strongly field-dependent strength:
as the field increases the spectral weight that is lost from the pole gets transferred to the two-magnon continuum.

In conclusion, we have analyzed the effect of a staggered magnetic field on a $S=1$ Haldane chain with single-ion anisotropy.
We have found that the zero-field splitting of the Haldane triplet due to the anisotropy term gets enhanced by the
presence of the field.
In addition, as the field is increased, the longitudinal mode has a spectral weight that decreases quite fast and is
transferred to a continuum, which has a two-magnon threshold.
We remark that this seems to be consistent with recent neutron scattering experiments on quasi-one-dimensional
compounds \cite{R}.

\section*{Acknowledgments}
We are grateful to C. Degli Esposti Boschi, F. Ortolani and P. Pieri for useful discussions during the preparation of this work.



\begin{thebibliography}{99}

\bibitem{DR} J. Darriet and L.P. regnault, Solid State Commun. {\bf 96}, 409 (1993);\\
J.F. DiTusa {\it et al}., Physica {\bf B194}, 181 (1994).

\bibitem{H} F.D.M. Haldane, Phys. Rev. Lett. {\bf 50}, 1153 (1983).

\bibitem{ZY}  A. Zheludev {\it et al}.,  Phys. Rev. \textbf{B54}, 6437 and 7210 (1996);\\
T. Yokoo {\it et al}.,  Phys. Rev. \textbf{B58}, 14424 (1998) and references therein.

\bibitem{Z}  A. Zheludev {\it et al}.,   Phys. Rev. Letters {\bf 80}, 3830 (1998).
\bibitem{R} S. Raymond  {\it et al}.,   Phys. Rev. Letters {\bf 82}, 2382 (1999).

\bibitem{MZ}  S. Maslov and A. Zheludev, Phys. Rev. Letters {\bf 80}, 5786 (1998).

\bibitem{A} I. Affleck, in: \textit{Fields, Strings and Critical Phenomena}, E. Brezin and J. Zinn-Justin Eds., North-Holland, Amsterdam, 1990.

\bibitem{Y} J. Lou, X. Dai, S. Qin, Z. Su and Yu Lu, Phys. Rev. \textbf{B60}, 52 (1999).

\bibitem{EMPR} E. Ercolessi, G. Morandi, P. Pieri and M. Roncaglia, Phys. Rev. B {\bf 62}, 14860 (2000) and Europhys. Lett. {\bf 52}, 434 (2000).

\bibitem{EMR} E. Ercolessi, G. Morandi  and M. Roncaglia, preprint cond-mat/0205234.

\bibitem{Cs} M. Kenzelmann et al.,  Phys. Rev. Letters {\bf 87}, 017201 (2001), cond-mat/011252.

\bibitem{O} F. Ortolani {\it et al}., in preparation.

\bibitem{HB} P.C. Hohenberg and W.F. Brinkman, Phys. Rev. B {\bf 10}, 128 (1974);\\
S.M. Girvin, A.H. Macdonald and P.M. Platzman, Phys. Rev. B {\bf 33}, 2481 (1986).



\end{thebibliography}
\end{document}